\documentclass[seceq]{ptptex}
\usepackage{wrapft}
\usepackage{graphicx}

\newcommand{\ket}[1]{| {#1} \rangle}     
\newcommand{\kket}[1]{| {#1} \rangle\!\rangle}     
\newcommand{\rket}[1]{| {#1} )}     
\newcommand{\dket}[1]{|\!| {#1} \rangle}     
\newcommand{\wtilde}[1]{\widetilde{#1}} 

\def\beq{\begin{eqnarray}}
\def\eeq{\end{eqnarray}}
\def\bsub{\begin{subequations}}
\def\esub{\end{subequations}}
\def\b{\begin{equation}}

\title{
Two-Level Lipkin Model in Unconventional Boson Realization
}

\author{
Yasuhiko {\sc Tsue},$^{1}$ 
Constan\c{c}a {\sc Provid\^encia},$^{2}$ 
Jo\~ao da {\sc Provid\^encia}$^{2}$ and 
Masatoshi {\sc Yamamura}$^{3}$  
}

\inst{
$^{1}$Physics Division, Faculty of Science, Kochi University, Kochi 780-8520, 
Japan\\
$^{2}$Departamento de F\'{i}sica, Universidade de Coimbra, 3004-516 Coimbra, 
Portugal\\
$^{3}$Department of Pure and Applied Physics, 
Faculty of Engineering Science,\\
Kansai University, Suita 564-8680, Japan
\\

}

\recdate{
\today
}

\abst{
With the aim of applying to the Lipkin model in the 
case of open shell system, a possible form of the boson realization for the 
$su(2)$-algebra is proposed both in the Schwinger and the Holstein-Primakoff representation. 
The basic idea is borrowed from the Schwinger boson realization for the 
$su(4)$-algebra for many-quark system recently presented by the present authors. 
As the simplest approximation, a certain approximate form is given and 
the result is a natural generalization of the RPA result in the 
case of closed shell system. 
}

\begin{document}

\maketitle

\section{Introduction}

It is well known that, with the aid of boson operators, we can describe 
various phenomena of nuclear and hadron physics successfully. 
Especially, the studies of microscopic structures of the boson operators 
trace back to the year 1960. 
In this year, Marumori, Arvieu \& Veneroni and Baranger\cite{1} proposed 
independently a theory, which is called as the quasi-particle random phase approximation. 
With the aid of this theory, we could understand microscopic structure of the boson operators 
describing the collective vibrational motion observed in the spherical nuclei. 
Further, the success of this theory has stimulated the studies of higher 
order corrections and one of the goal is the boson expansion theory: 
Belyaev \& Zelevinsky, Marumori, Yamamura \& Tokunaga, 
da Provid\^encia \& Weneser and Marshalek.\cite{2}
We can find various further studies concerning the boson expansion theory in the 
review by Klein \& Marshalek.\cite{3} 
Especially, this review concentrated on the boson realization of Lie algebra governing 
many-fermion system under consideration. 
The above is a rough sketch of the boson expansion theory 
at early stage. 
After these studies, too many papers have been published and it is impossible 
to follow them completely. 
Then, hereafter, we will narrow the discussion down to the boson realization 
of the $su(2)$-algebra.

We know two simple many-fermion models, which obey the $su(2)$-algebra: 
The single-level pairing model\cite{4} and the two-level Lipkin model.\cite{5} 
Each is composed of three $su(2)$-generators ${\wtilde S}_{\pm}$ and ${\wtilde S}_0$, 
which are of the bilinear forms for the fermion operators. 
Further, each contains the total fermion number operator ${\wtilde N}$. 
The operator ${\wtilde N}$ commutes with the Hamiltonian which is widely adopted. 
In the pairing model, ${\wtilde S}_0$ is a linear function of ${\wtilde N}$. 
Therefore, the change of the eigenvalue of ${\wtilde S}_0$ automatically 
leads to the change of the total fermion number. 
In the Lipkin model, ${\wtilde S}_0$ is proportional to the difference 
between the fermion number operators of the two levels. 
The total fermion number operator is the sum of both fermion number operators. 
Therefore, the change of the eigenvalue of ${\wtilde S}_0$ corresponds to 
the change of the difference between the fermion numbers of the two levels. 
The eigenvalues ${\wtilde S}_0$ and ${\wtilde N}$ are independent from each other. 
The above is an essential difference between the two models.

The single-level pairing model can be completely formulated in the 
frame of the conventional boson realization of the $su(2)$-algebra, namely, 
the Schwinger\cite{6} and the Holstein-Primakoff\cite{7} boson realization. 
In the former, the magnitude of the $su(2)$-spin is treated as $q$-number. 
In the later, it is treated as $c$-number. 
Of course, the total fermion number operator can be expressed in terms of the boson operators 
which are used for the generators. 
However, the case of the two-level Lipkin model is different from the case of the 
single-level pairing model. 
As was already mentioned, the total fermion number operator cannot be expressed in the frame of 
the boson operators in the conventional boson realization of the 
$su(2)$-algebra. 
From the above reason, usually, the total fermion number is fixed to the value 
corresponding to the closed shell system. 
Then, formally, we can apply the conventional boson realization of the $su(2)$-algebra 
used in the pairing model. 
Of course, the roles of various quantities appearing 
in the two models are different from each other. 
The above consideration suggests us that we must add a special device 
in order to complete the boson realization for the Lipkin model in the case 
of any fermion number.

Main aim of this paper is to present a possible form of the boson realization 
of the $su(2)$-algebra which is effective to the two-level Lipkin model 
in the case of any fermion number. 
We call it as unconventional boson realization, because conventional 
realization is not useful for the present case. 
The hint can be found in the Schwinger boson realization of the $su(4)$-algebra for 
the Bonn model and its modification describing many-quark system.\cite{8} 
The idea of these works is based on the Schwinger boson realization 
for the $su(4)$-algebra presented by the present author with 
Kuriyama and Kunihiro,\cite{9} which was 
intended to apply to the high temperature superconductivity.\cite{10}
The important conclusion is the following: 
In order to understand the quark-triplet formation as the important aspect 
of many-quark system, the symmetric representation is powerless. 
A certain non-symmetric representation, which is constructed 
under the extra degrees of freedom, should be adopted. 
Borrowing this idea, we can formulate the boson realization which makes our aim satisfy. 
We introduce the extra boson operators which do not contain in the conventional 
boson realization. 
The ideas in the $su(4)$- and the $su(2)$-algebra come from the Schwinger boson 
representation for the $su(M+1)\otimes su(N,1)$-algebra presented by the 
present authors with Kuriyama.\cite{11}
There exists a viewpoint that the Lipkin model was proposed as a model 
which enables us to describe particle-hole pair type 
collective excitation schematically, and therefore, 
it may be enough to investigate only the case of the closed shell system. 
However, it may be interesting to investigate how the collective 
excitation varies from the closed to the open shell system 
and it may be important to establish a theory, with 
the aid of which the above problem can be describe.

After recapitulating the two-level Lipkin model, in \S 2, we will 
discuss the conventional boson realization for both Schwinger and 
Holstein-Primakoff representation. 
In \S\S 3 and 4, we will formulate the Schwinger (\S 3) and 
the Holstein-Primakoff (\S 4) realization in unconventional form. 
In \S 5, the coupling of two $su(2)$-spins will be treated for 
the Lipkin model. 
Section 6 will be devoted to discussing the simplest approximation 
and it will be shown that the result is a natural generalization 
from that shown in the closed shell system based on RPA. 
In \S 7, the isosclar pairing model will be discussed. 
Finally, in \S 8, concluding remark will be given.

\section{Two-level Lipkin model and its conventional boson realizations}

Many-fermion system investigated in this paper is called the Lipkin model, 
which consists of two single-particle levels with the same degeneracy 
$2j+1\ (=2\Omega,\ j:$ half integer). 
We denote the two single-particle levels as the p- and the h-level. 
The fermion operators are denoted as $(c_{pm}^*,\ c_{pm})$ and 
$(c_{hm}^*,\ c_{hm})$, respectively, where $m=-j,\ -j+1, \cdots , 
j-1,\ j$. 
For the above system, we can define the following operators: 
\beq\label{2-1}
& &{\wtilde S}_+=\sum_{m}c_{pm}^*c_{hm} \ , \qquad
{\wtilde S}_-=\sum_{m}c_{hm}^*c_{pm} \ , \nonumber\\
& &{\wtilde S}_0=\frac{1}{2}\sum_m (c_{pm}^*c_{pm}-c_{hm}^*c_{hm}) \ . 
\eeq
The set $({\wtilde S}_{\pm,0})$ obeys the $su(2)$-algebra: 
\bsub\label{2-2}
\beq\label{2-2a}
& & [\ {\wtilde S}_+ \ , \ {\wtilde S}_-\ ]=2{\wtilde S}_0 \ , \qquad
[\ {\wtilde S}_0 \ , \ {\wtilde S}_{\pm} \ ]=\pm{\wtilde S}_{\pm} \ . 
\eeq
The Casimir operator ${\wtilde {\mib S}}^2$ is given in the form 
\beq\label{2-2b}
& &{\wtilde {\mib S}}^2={\wtilde S}_+{\wtilde S}_- 
+{\wtilde S}_0^2-{\wtilde S}_0 \ , 
\qquad
[\ {\wtilde {\mib S}}^2 \ , \ {\wtilde S}_{\pm,0} \ ]=0 \ .
\eeq
\esub
Further, the total fermion number operator ${\wtilde N}$ is given as 
\beq\label{2-3}
& & {\wtilde N}=\sum_m (c_{pm}^*c_{pm}+c_{hm}^*c_{hm}) \ , \qquad
[\ {\wtilde N} \ , \ {\wtilde S}_{\pm,0} \ ]=0 \ .
\eeq
It should be noted that ${\wtilde {\mib S}}^2$ and ${\wtilde N}$ commute 
with ${\wtilde S}_{\pm,0}$, but, they are independent of each other. 
Therefore, an orthogonal set is specified by the eigenvalues of 
${\wtilde N}$, ${\wtilde {\mib S}}^2$ and ${\wtilde S}_0$ for a given value 
of $\Omega$. 
This point is different from the single-level pairing model. 
The Hamiltonian of the present model ${\wtilde H}$ is usually expressed in 
the form 
\beq\label{2-4}
& &{\wtilde H}=\epsilon {\wtilde S}_0-\chi({\wtilde S}_+^2 + 
{\wtilde S}_-^2) \ . \quad 
(\epsilon \ , \ \chi > 0)
\eeq
Since $\epsilon >0$, energetically the p-level is higher than the h-level 
with the energy difference $\epsilon$. 
The parameter $\chi$ denotes the force strength. 
We also notice the relations 
$[{\wtilde N}\ , \ {\wtilde H}]=[{\wtilde {\mib S}}^2 \ , \ {\wtilde H}]=0$, 
but, $[{\wtilde S}_0 \ , \ {\wtilde H}] \neq 0$.

It may be convenient to express the above operators in terms of the 
particle and the hole operator: 
\beq\label{2-5}
& &c_{pm}=a_m \ , \qquad c_{hm}=b_{\wtilde m}^*\ (=(-)^{j-m}b_{-m}^*) \ .
\eeq
Then, ${\wtilde S}_{\pm,0}$ and ${\wtilde N}$ can be expressed as 
\beq
& &{\wtilde S}_+=\sum_m a_m^*b_{\wtilde m}^* \ , \qquad
{\wtilde S}_-=\sum_m b_{\wtilde m}a_m \ , \nonumber\\
& &{\wtilde S}_0=\frac{1}{2}\sum_m (a_m^* a_m+b_m^* b_m)-\Omega  \ , 
\label{2-6}\\
& &{\wtilde N}=\sum_m (a_m^* a_m - b_m^* b_m)+2\Omega \ . 
\label{2-7}
\eeq
The forms (\ref{2-6}) and (\ref{2-7}) give us 
\beq\label{2-8}
& &{\wtilde S}_0=\frac{1}{2}({\wtilde N}_p+{\wtilde N}_h)-\Omega \ , \qquad
{\wtilde N}={\wtilde N}_p-{\wtilde N}_h +2\Omega \ , 
\eeq
conversely, 
\beq\label{2-9}
& &{\wtilde N}_p={\wtilde S}_0+\frac{1}{2}{\wtilde N} \ , \qquad
{\wtilde N}_h={\wtilde S}_0-\frac{1}{2}{\wtilde N}+2\Omega\ . 
\eeq
Here, ${\wtilde N}_p$ and ${\wtilde N}_h$ denote the particle and the hole 
number operator, respectively: 
\beq\label{2-10}
& &{\wtilde N}_p=\sum_m a_m^*a_m \ , \qquad
{\wtilde N}_h=\sum_m b_m^*b_m \ . 
\eeq
The form (\ref{2-6}) tells us that ${\wtilde S}_+$ plays a role of the 
creation of the particle-hole pair coupled to the angular momentum 
$J=0$. 
Further, it should be noted that in contrast with the single-level 
pairing model, the total fermion number is not the simple sum 
of ${\wtilde N}_p$ and ${\wtilde N}_h$, namely ${\wtilde N}_p+{\wtilde N}_h$. 
This fact is in an important position in the present model.

The minimum weight state, which we denote as $\rket{m}$, is introduced 
as a state satisfying the condition 
\bsub\label{2-11}
\beq
& &{\wtilde S}_-\rket{m}=0 \ , \qquad
{\wtilde S}_0\rket{m}=-s\rket{m} \ , 
\label{2-11a}\\
& &{\wtilde N}\rket{m}=N\rket{m} \ . 
\label{2-11b}
\eeq
\esub
Therefore, the eigenstate of ${\wtilde S}_0$ with the eigenvalue $s_0$ 
can be expressed in the form 
\beq\label{2-12}
& &\rket{\Omega, N;ss_0}=({\wtilde S}_+)^{s+s_0}
\rket{\Omega,N;s} \ . 
\quad
\left( \rket{m}=\rket{\Omega,N;s} \right)
\eeq
The quantity $\Omega$ is not quantum number, but, the parameter 
characterizing the model. 
However, in our boson realization later we will discuss, $\Omega$ is 
treated as the quantum number. 
Then, we add $\Omega$ explicitly to the fermion states. 
It may be convenient to formulate the above result in terms of 
$n_p$ and $n_h$ which satisfy 
\beq\label{2-13}
& &{\wtilde N}_p\rket{m}=n_p\rket{m}\ , \qquad
{\wtilde N}_h\rket{m}=n_h\rket{m} \ . 
\eeq
The eigenvalue equation (\ref{2-13}) is permitted from the relation (\ref{2-9}). 
The relation (\ref{2-8}) gives 
\bsub\label{2-14}
\beq
& &s=\Omega-\frac{1}{2}(n_p+n_h) \ , 
\label{2-14a}\\
& &N=2\Omega +(n_p-n_h) \ . 
\label{2-14b}
\eeq
\esub
For the case $n_p=n_h\ (=n_0)$, we have 
\beq\label{2-15}
& &s=\Omega-n_0 \ , \qquad N=2\Omega\ . 
\eeq

The relation $N=2\Omega$ shows us that the case $n_p=n_h\ (=n_0)$ 
corresponds to the case of the closed shell, where if no residual 
interaction, the h-level is completely occupied by the fermions 
in the ground state. 
The quantity $n_0$ denotes the number of the particle-hole pairs 
with the coupled angular momentum $J\neq 0$. 
Conventionally, the two-level Lipkin model has been investigated in 
this case. 
For this case, the Schwinger boson realization of the $su(2)$-algebra is 
applicable in the following form: 
\beq\label{2-16}
& &{\wtilde S}_+ \rightarrow {\hat S}_+={\hat a}^*{\hat b}\ , \quad
{\wtilde S}_- \rightarrow {\hat S}_-={\hat b}^*{\hat a}\ , \quad
{\wtilde S}_0 \rightarrow {\hat S}_0=\frac{1}{2}({\hat a}^*{\hat a}
-{\hat b}^*{\hat b})\ . 
\eeq
Here, $({\hat a},{\hat a}^*)$ and $({\hat b},{\hat b}^*)$ denote two kinds 
of boson operators. 
In this case, the state $\rket{\Omega,N=2\Omega;ss_0}$ corresponds to 
\beq\label{2-17}
\rket{\Omega,N=2\Omega;ss_0} & & \nonumber\\
\rightarrow \ \ket{\Omega,N=2\Omega;ss_0}&=&
({\hat a}^*{\hat b})^{s+s_0}({\hat b}^*)^{2s}\ket{0} \nonumber\\
&=& ({\hat a}^*)^{s+s_0}({\hat b}^*)^{s-s_0}\ket{0}\ , 
\qquad
\ket{m}=({\hat b}^*)^{2s}\ket{0} \ .
\eeq
Here, $\ket{0}$ denotes the vacuum of the bosons and, of course, 
$s$ is given in the relation (\ref{2-15}). 
With the aid of the relation (\ref{2-15}), we can introduce the operator 
${\hat \Omega}$ expressing the degeneracy of the single-particle 
levels. 
The operator ${\hat \Omega}$ commutes with $({\hat S}_{\pm,0})$, 
because the eigenvalue of ${\hat \Omega}$ should not depend on 
individual eigenstates. 
Then, we may set up ${\hat \Omega}$ in the form 
\beq\label{2-18}
& &{\hat \Omega}=x+\frac{1}{2}y({\hat a}^*{\hat a}+{\hat b}^*{\hat b}) \ .
\eeq
Here, $x$ and $y$ denote constants to be determined. 
Operation of ${\hat \Omega}$ on $\ket{m}$ given in the relation 
(\ref{2-17}) leads to 
\beq\label{2-19}
& &\Omega=x+ys=x+y(\Omega-n_0) \ , \quad
{\rm i.e.,}\quad 
x=(1-y)\Omega +yn_0 \ .
\eeq
Substituting the form (\ref{2-19}) into the relation (\ref{2-18}), we have 
\beq\label{2-20}
& &{\hat \Omega}=(1-y)\Omega+y\left(
n_0+\frac{1}{2}({\hat a}^*{\hat a}+{\hat b}^*{\hat b})\right) \ .
\eeq
It is not natural that ${\hat \Omega}$ should depend on the eigenvalue. 
Therefore, we set $y=1$, which leads to 
\beq\label{2-21}
& &{\hat \Omega}=n_0+\frac{1}{2}({\hat a}^*{\hat a}+{\hat b}^*{\hat b}) \ .
\eeq
The relation (\ref{2-21}) shows that $n_0$ is a parameter, the value of 
which should be given from the outside. 
Total fermion number operator ${\hat N}$ is given as 
\beq\label{2-22}
& &{\hat N}=2{\hat \Omega} \ .
\eeq
The above is the Schwinger boson realization. 
We know another boson realization called the Holstein-Primakoff 
boson realization. 
With the use of one kind of boson $({\hat A},{\hat A}^*)$, 
${\wtilde S}_{\pm,0}$ corresponds to ${\hat S}_{\pm,0}(s)$, which 
is given in the form
\beq\label{2-23}
& &{\wtilde S}_+ \rightarrow 
{\hat S}_+(s)={\hat A}^*\sqrt{2s-{\hat A}^*{\hat A}} \ , \qquad
{\wtilde S}_- \rightarrow 
{\hat S}_-(s)=\sqrt{2s-{\hat A}^*{\hat A}} \ {\hat A} \ , \nonumber\\
& &{\wtilde S}_0 \rightarrow 
{\hat S}_0(s)={\hat A}^*{\hat A}-s \ . 
\eeq
The state $\rket{\Omega,N=2\Omega;ss_0}$ corresponds to 
\beq\label{2-24}
\rket{\Omega,N=2\Omega;ss_0} & &\nonumber\\
\rightarrow \kket{\Omega,N=2\Omega;ss_0}&=&
\left({\hat A}^*\sqrt{2s-{\hat A}^*{\hat A}}\right)^{s+s_0}\kket{s} \nonumber\\
&=&\left({\hat A}^*\right)^{s+s_0}\kket{s} \ , 
\qquad \kket{m}=\kket{s}\ . 
\eeq
Here, $\kket{s}$ denotes the boson vacuum satisfying 
${\hat A}\kket{s}=0$ and ${\hat S}_0(s)\kket{s}=-s\kket{s}$ is derived from 
the relation (\ref{2-23}). 
The relation between both realizations will be discussed in \S 4.

From the above argument, we could learn that the conventional boson 
representations of the $su(2)$-algebra are only applicable to the case 
$N=2\Omega$ in the two-level Lipkin model. 
This conclusion induces a problem how to treat the case 
$n_p\neq n_h$. 
The relation (\ref{2-14}) teaches us the following: 
\bsub\label{2-25}
\beq
& &{\rm if}\ \ n_p < n_h \ , \quad N < 2\Omega \ , 
\label{2-25a}\\
& &{\rm if}\ \ n_p > n_h \ , \quad N > 2\Omega \ . 
\label{2-25b}
\eeq
\esub
The case (\ref{2-25a}) corresponds to the system in which $n_p$ particle-hole 
pairs with $J\neq 0$ exist and $(n_h-n_p)$ holes cannot couple with 
any particle. 
In the case (\ref{2-25b}), $n_h$ particle-hole pairs with $J\neq 0$ 
exist and $(n_p-n_h)$ particles cannot couple with any hole. 
The relation (\ref{2-14}) is summarized as follows: \\
(i) The case $n_p \leq n_h$: 
\bsub\label{2-26}
\beq\label{2-26a}
& &N=2\Omega-(n_h-n_p) \ , \qquad s=\frac{1}{2}N-n_p \ . 
\eeq
Since $n_h-n_p \geq 0$, $N \geq 0$ and $s\geq 0$, the relation 
(\ref{2-26a}) gives us the inequalities 
\beq\label{2-26b}
& &0 \leq N \leq 2\Omega \ , \qquad n_p\leq \frac{1}{2}N \ , \qquad
n_h-n_p \leq 2\Omega \ . 
\eeq
\esub
(ii) The case $n_p \geq n_h$: 
\bsub\label{2-27}
\beq\label{2-27a}
& &N=2\Omega+(n_p-n_h) \ , \qquad s=\frac{1}{2}(4\Omega-N)-n_h \ . 
\eeq
Since $n_p-n_h \geq 0$, $N \leq 4\Omega$ and $s\geq 0$, the relation 
(\ref{2-27a}) gives us the inequalities 
\beq\label{2-27b}
& &2\Omega \leq N \leq 4\Omega \ , \qquad n_h\leq \frac{1}{2}(4\Omega-N) 
\ , \qquad
n_p-n_h \leq 2\Omega \ . 
\eeq
\esub
In the next section, we discuss the boson realization including 
$n_p \neq n_h$.

\section{Unconventional boson realization : Part (I)}

The discussion in \S 2 suggests us that, in order to make the boson 
realization of the $su(2)$-algebra applicable to the Lipkin model 
including the case $n_p\neq n_h$, we have to introduce extra degrees 
of freedom. 
For this purpose, we apply a form proposed by the present authors, 
in which the Schwinger boson representation of the $su(M+1)$-algebra 
is formulated in terms of $(M+1)(N+1)$ kinds of bosons.\cite{11} 
The case $(M=3,N=1)$ has been applied to the Bonn model 
and its modification for many-quark system.\cite{8} 
This model obeys the $su(4)$-algebra. 
The case $(M=1,N=0)$ corresponds to the $su(2)$-algebra adopted in \S 2. 
In this section, we treat the case $(M=1,N=1)$, which contains four 
kinds of bosons. 
Under the notations appropriate to the present case, 
${\hat S}_{\pm,0}$ can be expressed in the form 
\beq\label{3-1}
& &{\hat S}_+={\hat a}_p^*{\hat b}_h-{\hat a}_h^*{\hat b}_p \ , \qquad
{\hat S}_-={\hat b}_h^*{\hat a}_p-{\hat b}_p^*{\hat a}_h \ , \nonumber\\
& &{\hat S}_0=\frac{1}{2}\left[
({\hat a}_p^*{\hat a}_p+{\hat a}_h^*{\hat a}_h)
-({\hat b}_p^*{\hat b}_p+{\hat b}_h^*{\hat b}_h)\right] \ .
\eeq
Here, $({\hat a}_p,{\hat a}_p^*)$, $({\hat a}_h,{\hat a}_h^*)$, 
$({\hat b}_p,{\hat b}_p^*)$ and $({\hat b}_h,{\hat b}_h^*)$ 
denote four kinds of bosons. 
Associating with the $su(2)$-algebra, we can define the $su(1,1)$-algebra 
in the form 
\beq\label{3-2}
& &{\hat T}_+={\hat a}_p^*{\hat b}_p^* +{\hat a}_h^*{\hat b}_h^* \ , \qquad
{\hat T}_-={\hat b}_p{\hat a}_p+{\hat b}_h{\hat a}_h \ , \nonumber\\
& &{\hat T}_0=\frac{1}{2}\left[
({\hat a}_p^*{\hat a}_p+{\hat a}_h^*{\hat a}_h)
+({\hat b}_p^*{\hat b}_p+{\hat b}_h^*{\hat b}_h)\right]+1 \ .
\eeq
The set $({\hat T}_{\pm,0})$ obeys 
\beq
& &[\ {\hat T}_+ \ , \ {\hat T}_-\ ]=-2{\hat T}_0 \ , \qquad
[\ {\hat T}_0 \ , \ {\hat T}_{\pm}\ ]=\pm{\hat T}_{\pm} \ , 
\label{3-3}\\
& &[\ {\rm any\ of}\ ({\hat T}_{\pm,0}) \ , 
\ {\rm any\ of}\ ({\hat S}_{\pm,0})\ ]=0 \ . 
\label{3-4}
\eeq
The Casimir operator ${\hat {\mib T}}^2$ is given as 
\beq\label{3-5}
& &{\hat {\mib T}}^2=-{\hat T}_+{\hat T}_- + {\hat T}_0^2+{\hat T}_0 \ .
\eeq
It is noted that the operator ${\hat M}$ defined in the following commutes 
with any of $({\hat S}_{\pm,0})$ and $({\hat T}_{\pm,0})$: 
\beq
& &{\hat M}=({\hat a}_p^*{\hat a}_p-{\hat b}_p^*{\hat b}_p)
-({\hat a}_h^*{\hat a}_h-{\hat b}_h^*{\hat b}_h)\ , 
\label{3-6}\\
& &[\ {\hat M}\ , \ {\hat S}_{\pm,0}\ ]=[\ {\hat M}\ , \ {\hat T}_{\pm,0}\ ]
=0 \ .
\label{3-7}
\eeq
The above is an outline of the Schwinger boson representation of the 
$su(2)$-algebra in terms of four kinds of bosons. 
For this form, we must pay an attention to the following: 
The above boson representation is not the boson realization of the Lipkin 
model as it stands, because the above does not contain the 
degeneracy operator ${\hat \Omega}$ and the total fermion number 
operator ${\hat N}$ which connect with the original many-fermion 
system.

As for the minimum weight state $\ket{m}$, which corresponds to 
$\rket{m}$, we postulate the following state: 
\beq\label{3-8}
& &\ket{m}=({\hat b}_p^*)^{|n_p-n_h|}({\hat b}_h^*)^{2\Omega-
(n_p+n_h)-|n_p-n_h|}\ket{0} \ .
\eeq
Clearly, $\ket{m}$ satisfies 
\beq\label{3-9}
& &{\hat S}_-\ket{m}=0 \ , \qquad
{\hat S}_0\ket{m}=-s\ket{m} \ , \qquad
s=\Omega-\frac{1}{2}(n_p+n_h) \ . 
\eeq
The relation (\ref{3-9}) corresponds to the relation (\ref{2-11a}) with 
(\ref{2-14a}). 
If $n_p=n_h\ (=n_0)$, the state (\ref{3-8}) reduces to 
$\ket{m}=({\hat b}_h^*)^{2s}\ket{0}\ (s=\Omega-n_0)$ and the $(s+s_0)$-time 
operation of ${\hat S}_+$ on $\ket{m}$ gives us 
$({\hat a}_p^*)^{s+s_0}({\hat b}_h^*)^{s-s_0}\ket{0}$ in the space spanned 
by four kinds of bosons. 
If ${\hat a}_p$ and ${\hat b}_h$ read ${\hat a}$ and ${\hat b}$, respectively, 
the above form reduces to the form (\ref{2-17}). 
The above argument supports that $\ket{m}$ defined in the relation (\ref{3-8}) 
may be regarded as the minimum weight state for our purpose. 
However, we must notice the connection of the state (\ref{3-8}) to the 
$su(1,1)$-algebra (\ref{3-2}). 
The state $\ket{m}$ satisfies the relation: 
\beq\label{3-10}
& &{\hat T}_-\ket{m}=0\ , \qquad
{\hat T}_0\ket{m}=(s+1)\ket{m}\ , \qquad
s=\Omega-\frac{1}{2}(n_p+n_h)\ .
\eeq
Next, we introduce the following state: 
\beq\label{3-11}
& &\ket{n;m}=({\hat T}_+)^n\ket{m} \ , \quad (n=0,\ 1,\ 2,\cdots) \nonumber\\
& &\ket{0;m}=\ket{m}\ . 
\eeq
The state (\ref{3-11}) leads us to the same form as the relation 
(\ref{3-8}): 
\beq\label{3-12}
& &{\hat S}_-\ket{n;m}=0 \ , \qquad
{\hat S}_0\ket{n;m}=-s\ket{n;m}\ , \qquad 
s=\Omega-\frac{1}{2}(n_p+n_h) \ .
\eeq
The above indicates that the present boson space is not in 
one-to-one correspondence with the original fermion space. 
Then, in order to guarantee the one-to-one correspondence, we require the 
condition that our minimum weight state is also the minimum weight state 
for the $su(1,1)$-algebra. 
The state $\ket{m}$ satisfies this condition: 
\beq\label{3-13}
& &{\hat T}_-\ket{m}=0\ , \qquad
{\hat T}_0\ket{m}=(s+1)\ket{m}\ , \qquad
s=\Omega-\frac{1}{2}(n_p+n_h) \ . 
\eeq
Then, if $\Omega$, $n_p$ and $n_h$ are specified from the outside, 
$\ket{m}$ is given in the form (\ref{3-8}), which corresponds to $\rket{m}$. 
Further, if the total fermion number is given as $N=2\Omega+n_p-n_h$, 
the expression $s=\Omega-(n_p+n_h)/2$ gives us $\ket{m}$ 
in the form $\ket{\Omega,N;s}$ which corresponds to the state 
$\rket{\Omega,N;s}$ shown in the relation (\ref{2-12}). 
The orthogonal state $\ket{\Omega,N;ss_0}$ is given in the form 
\beq\label{3-14}
& &\ket{\Omega,N;ss_0}=({\hat S}_+)^{s+s_0}\ket{\Omega,N;s}\ .
\eeq
In \S 4, we will repeat the above discussion in a way slightly 
different from the above.

Our final task is to search ${\hat \Omega}$ and ${\hat N}$ which correspond 
to the degeneracy of the single-particle level $\Omega$ and the total 
fermion number operator ${\wtilde N}$. 
If we cannot search these two operators, our boson representation would not be 
permitted to call the boson realization of the Lipkin model. 
First, we treat ${\hat \Omega}$. 
Under the same viewpoint as that in \S 2, we set up the following form: 
\beq\label{3-15}
& &{\hat \Omega}=x+\frac{y}{2}
({\hat a}_p^*{\hat a}_p+{\hat a}_h^*{\hat a}_h+{\hat b}_p^*{\hat b}_p 
+{\hat b}_h^*{\hat b}_h) \ .
\eeq
With the aid of the relation ${\hat \Omega}\ket{m}=\Omega\ket{m}$, 
we obtain 
\beq\label{3-16}
& &x=(1-y)\Omega+\frac{y}{2}(n_p+n_h) \ . 
\eeq
Then, ${\hat \Omega}$ is expressed as the form 
\beq\label{3-17}
& &{\hat \Omega}=
(1-y)\Omega+\frac{y}{2}(n_p+n_h+
{\hat a}_p^*{\hat a}_p+{\hat a}_h^*{\hat a}_h+{\hat b}_p^*{\hat b}_p 
+{\hat b}_h^*{\hat b}_h) \ .
\eeq
In the same idea as that adopted in \S 2, we have $y=1$ and ${\hat \Omega}$ 
is expressed as
\beq\label{3-18}
& &{\hat \Omega}=
\frac{1}{2}(n_p+n_h)+\frac{1}{2}
({\hat a}_p^*{\hat a}_p+{\hat a}_h^*{\hat a}_h+{\hat b}_p^*{\hat b}_p 
+{\hat b}_h^*{\hat b}_h) \ .
\eeq
The above is a natural generalization of the form (\ref{2-21}).

Our next task is to find the operator ${\hat N}$. 
For this aim, first, we pay an attention to the operators 
${\wtilde N}_p$ and ${\wtilde N}_h$ in the form (\ref{2-10}). 
The operators ${\wtilde N}_p$ and ${\wtilde N}_h$ have the following 
properties: 
\beq
& &{\wtilde N}_p ,\ {\wtilde N}_h \ : \ {\rm positive\ definite}, 
\label{3-19}\\
& &[\ {\wtilde N}_p\ , \ {\wtilde S}_{\pm}\ ]=\pm{\wtilde S}_{\pm} \ ,\qquad
[\ {\wtilde N}_h\ , \ {\wtilde S}_{\pm}\ ]=\pm{\wtilde S}_{\pm} \ ,
\label{3-20}\\
& &{\wtilde N}_p\rket{m}=n_p\rket{m}\ , \qquad
{\wtilde N}_h\rket{m}=n_h\rket{m} \ . 
\label{3-21}
\eeq
The relation (\ref{3-21}) is a copy from the relation (\ref{2-13}). 
We search ${\hat N}_p$ and ${\hat N}_h$ satisfying the relations which 
correspond to the relations (\ref{3-19})$\sim$(\ref{3-21}). 
For this purpose, we define ${\hat N}_p$ and ${\hat N}_h$ in the form 
\bsub\label{3-22}
\beq
& &{\hat N}_p={\hat M}_p+x_p+y_p{\hat \Omega}\ , \qquad
{\hat M}_p={\hat a}_p^*{\hat a}_p-{\hat b}_p^*{\hat b}_p \ , 
\label{3-22a}\\
& &{\hat N}_h={\hat M}_h+x_h+y_h{\hat \Omega}\ , \qquad
{\hat M}_h={\hat a}_h^*{\hat a}_h-{\hat b}_h^*{\hat b}_h \ . 
\label{3-22b}
\eeq
\esub
Here, $(x_p,y_p)$ and $(x_h,y_h)$ denote parameters to be determined. 
It should be noted that ${\hat M}_p$ and ${\hat M}_h$ satisfy 
\beq
& &[\ {\hat M}_p \ , \ {\hat S}_{\pm}\ ]=\pm{\hat S}_{\pm} \ , \qquad
[\ {\hat M}_h \ , \ {\hat S}_{\pm}\ ]=\pm{\hat S}_{\pm} \ , 
\label{3-23}\\
& &[\ {\hat M}_p \ , \ {\hat T}_{\pm}\ ]=0 \ , \qquad
[\ {\hat M}_h \ , \ {\hat T}_{\pm}\ ]=0 \ . 
\label{3-24}
\eeq
The operators ${\hat M}_p$ and ${\hat M}_h$ satisfy the same relation as 
that shown in the relation (\ref{3-20}), but they are not positive definite. 
Then, we add the terms $(x_p+y_p{\hat \Omega})$ and 
$(x_h+y_h{\hat \Omega})$ which commute with ${\hat S}_{\pm}$. 
By using the relations ${\hat N}_p\ket{m}=n_p\ket{m}$ and 
${\hat N}_h\ket{m}=n_h\ket{m}$ which correspond to the relation (\ref{3-21}), 
we have the relation 
\beq\label{3-25}
& &(x_p-n_p-|n_p-n_h|)+y_p\Omega=0 \ , \nonumber\\
& &(x_h+n_p+|n_p-n_h|)+(y_h-2)\Omega=0 \ . 
\eeq
For the relation (\ref{3-25}), we require that $(x_p,y_p)$ and 
$(x_h,y_h)$ should not depend on $\Omega$ which is an eigenvalue of 
${\hat \Omega}$. 
Under this requirement, we have 
\beq\label{3-26}
& &x_p=n_p+|n_p-n_h| \ , \qquad y_p=0 \ , \nonumber\\
& &x_h=-n_p-|n_p-n_h| \ , \qquad
y_h=2 \ . 
\eeq
Under the relation (\ref{3-20}) and the explicit expressions of 
${\hat \Omega}$, ${\hat M}_p$ and ${\hat M}_h$ shown in the relations 
(\ref{3-18}), (\ref{3-22a}) and (\ref{3-22b}), we have 
\beq\label{3-27}
& &{\hat N}_p=n_p+|n_p-n_h|+{\hat a}_p^*{\hat a}_p-{\hat b}_p^*{\hat b}_p \ , 
\nonumber\\
& &{\hat N}_h=n_h-|n_p-n_h|+{\hat a}_p^*{\hat a}_p
+2{\hat a}_h^*{\hat a}_h+{\hat b}_p^*{\hat b}_p \ .
\eeq
Since ${\hat N}_p({\hat S}_+)^{n}\ket{m}=(n+n_p)({\hat S}_+)^{n}\ket{m}$ 
and ${\hat N}_h({\hat S}_+)^{n}\ket{m}=(n+n_h)({\hat S}_+)^{n}\ket{m}$, 
${\hat N}_p$ and ${\hat N}_h$ are positive definite. 
Under the correspondence to the relation (\ref{2-7}) for ${\wtilde N}$, 
we define ${\hat N}$ in the form 
\beq\label{3-28}
& &{\hat N}={\hat N}_p-{\hat N}_h+2{\hat \Omega} \ . 
\eeq
Substituting the relations (\ref{3-18}) and (\ref{3-27}) into the definition 
(\ref{3-28}), we have 
\beq\label{3-29}
& &{\hat N}=2(n_p+|n_p-n_h|)
+({\hat a}_p^*{\hat a}_p-{\hat a}_h^*{\hat a}_h-
{\hat b}_p^*{\hat b}_p+{\hat b}_h^*{\hat b}_h) \ .
\eeq
With the use of ${\hat M}$ defined in the relation (\ref{3-6}), 
${\hat N}$ can be given as 
\beq\label{3-30}
{\hat N}=2(n_p+|n_p-n_h|)+{\hat M}\ . 
\eeq
The relation (\ref{3-7}) leads to 
\beq\label{3-31}
[\ {\hat N} \ , \ {\hat S}_{\pm,0}\ ]=[\ {\hat N}\ , \ {\hat T}_{\pm,0}\ ]
=0 \ .
\eeq

Thus, we could formulate an unconventional boson realization for the 
Lipkin model including the case $n_p\neq n_h$.

\section{Unconventional boson realization : Part (II)}

Main task of this section is to formulate the Lipkin model including the 
case $n_p\neq n_h$ in the frame of the Holstein-Primakoff representation 
derived in relation to the Schwinger boson representation. 
For the preparation, first, we treat the case $n_p=n_h\ (= n_0)$ 
discussed in \S 2. 
The interpretation is suggestive, but not strict. 
The strict interpretation has been given 
in Ref.\citen{12}. 
In the present case, the operator introduced in the relation 
(\ref{2-21}), ${\hat \Omega}$, plays an essential role: 
\beq\label{4-1}
& &{\hat \Omega}=n_0+{\hat S} \ , \qquad
{\hat S}=\frac{1}{2}({\hat a}^*{\hat a}+{\hat b}^*{\hat b}) \ .
\eeq
The operator ${\hat S}$ commutes with ${\hat S}_{\pm,0}$. 
Therefore, the irreducible representation is specified by the 
eigenvalue of ${\hat S}$ which we denote $s$. 
In the space specified by $s$, it may be permitted to set up
\beq\label{4-2}
& &\frac{1}{2}({\hat a}^*{\hat a}+{\hat b}^*{\hat b})=s\ , \ \ 
{\rm i.e.,}\ \ {\hat b}^*{\hat b}=2s-{\hat a}^*{\hat a}
=\left(\sqrt{2s-{\hat a}^*{\hat a}}\right)^2 \ . 
\eeq
Of course, we have 
\beq\label{4-3}
& &{\hat S}\rightarrow s=\Omega-n_0 \ .
\eeq
Therefore, it should be noted that even if $({\hat a},{\hat a}^*)$ is boson, 
$({\hat b},{\hat b}^*)$ cannot be regarded as boson independent 
of $({\hat a},{\hat a}^*)$. 
Under the above consideration, we set up the relation (\ref{2-23}).

Under the idea mentioned above, we will present the Holstein-Primakoff 
representation including the case $n_p\neq n_h$. 
As the operators playing the same role as ${\hat S}$ given in the 
relation (\ref{4-1}), we introduce the following two operators: 
\bsub\label{4-4}
\beq
& &{\hat S}_p=\frac{1}{2}({\hat a}_p^*{\hat a}_p+{\hat b}_h^*{\hat b}_h) \ , 
\label{4-4a}\\
& &{\hat S}_h=\frac{1}{2}({\hat a}_h^*{\hat a}_h+{\hat b}_p^*{\hat b}_p) \ . 
\label{4-4b}
\eeq
\esub
They satisfy the relation 
\beq\label{4-5}
& &[\ {\hat S}_p \ , \ {\hat S}_{\pm,0}\ ]
=[\ {\hat S}_h\ , \ {\hat S}_{\pm,0}\ ]=0 \ . 
\eeq
Therefore, the irreducible representation is specified by the eigenvalues of 
${\hat S}_p$ and ${\hat S}_h$, which are denoted by $s_p$ and $s_h$, 
respectively. 
The relations (\ref{3-18}) and (\ref{3-29}) lead to 
\bsub\label{4-6}
\beq
& &{\hat S}_p=\frac{1}{2}\left({\hat \Omega}+\frac{1}{2}{\hat N}\right)
-\frac{1}{4}(3n_p+n_h)-\frac{1}{2}|n_p-n_h| \ , 
\label{4-6a}\\
& &{\hat S}_h=\frac{1}{2}\left({\hat \Omega}-\frac{1}{2}{\hat N}\right)
+\frac{1}{4}(n_p-n_h)+\frac{1}{2}|n_p-n_h| \ . 
\label{4-6b}
\eeq
\esub
Therefore, we have 
\bsub\label{4-7}
\beq
{\hat S}_p \rightarrow 
s_p&=&\frac{1}{2}\left({\Omega}+\frac{1}{2}{N}\right)
-\frac{1}{4}(3n_p+n_h)-\frac{1}{2}|n_p-n_h| \nonumber\\
&=&\Omega-\frac{1}{2}(n_p+n_h)-\frac{1}{2}|n_p-n_h| \ , 
\label{4-7a}\\
{\hat S}_h \rightarrow 
s_h&=&\frac{1}{2}\left({\Omega}-\frac{1}{2}{N}\right)
+\frac{1}{4}(n_p-n_h)+\frac{1}{2}|n_p-n_h| \nonumber\\
&=&\frac{1}{2}|n_p-n_h| \ . 
\label{4-7b}
\eeq
\esub
Here, we used the relation (\ref{3-28}): 
\beq\label{4-8}
& &{\hat N} \rightarrow N=n_p-n_h+\Omega \ . 
\eeq
The relation (\ref{4-7}) corresponds to the relation (\ref{4-3}). 
The relation (\ref{4-7}) gives us 
\beq\label{4-9}
& &s_p+s_h=\Omega-\frac{1}{2}(n_p+n_h)=s\ . 
\eeq
Here, we used the relation (\ref{2-14a}).

In the same idea as the previous case, we set up 
the following relation which comes from the relation (\ref{4-4}): 
\bsub\label{4-10}
\beq
& &{\hat b}_h^*{\hat b}_h=2s_p-{\hat a}_p^*{\hat a}_p
=\left(\sqrt{2s_p-{\hat a}_p^*{\hat a}_p}\right)^2 \ , 
\label{4-10a}\\
& &{\hat b}_p^*{\hat b}_p=2s_h-{\hat a}_h^*{\hat a}_h
=\left(\sqrt{2s_h-{\hat a}_h^*{\hat a}_h}\right)^2 \ . 
\label{4-10b}
\eeq
\esub
The form (\ref{3-1}) gives us the idea for the relation 
\beq\label{4-11}
& &{\wtilde S}_+ \rightarrow 
{\hat S}_+(s_p s_h)={\hat A}_p^*\sqrt{2s_p-{\hat A}_p^*{\hat A}_p}
-{\hat A}_h^*\sqrt{2s_h-{\hat A}_h^*{\hat A}_h} \ , \nonumber\\
& &{\wtilde S}_- \rightarrow 
{\hat S}_-(s_p s_h)=\sqrt{2s_p-{\hat A}_p^*{\hat A}_p}\ {\hat A}_p
-\sqrt{2s_h-{\hat A}_h^*{\hat A}_h}\ {\hat A}_h \ , \nonumber\\
& &{\wtilde S}_0 \rightarrow 
{\hat S}_0(s_p s_h)={\hat A}_p^*{\hat A}_p+
{\hat A}_h^*{\hat A}_h-s \ .
\eeq
Here, we used the relation (\ref{4-9}). 
The form (\ref{4-1}) is nothing but the Holstein-Primakoff representation. 
If $n_p=n_h\ (=n_0)$, $s_h$ vanishes and $s_p=s$. 
Then, if ${\hat A}_p$ reads ${\hat A}$, 
the form (\ref{4-11}) is reduced to the form 
(\ref{2-23}). 
This is in the same situation as the case of the Schwinger representation. 
The minimum weight state in the present case, 
which we denote as $\dket{m}$, is given as the vacuum of the 
bosons ${\hat A}_p$ and ${\hat A}_h$: 
\beq
& &{\hat A}_p\dket{m}={\hat A}_h\dket{m}=0 \ , 
\label{4-12}\\
{\rm i.e.,}\ \ 
& &{\hat S}_-(s_ps_h)\dket{m}=0 \ , \qquad
{\hat S}_0(s_ps_h)\dket{m}=-s\dket{m}\ . 
\label{4-13}
\eeq
Since $s=\Omega-(n_p+n_h)/2$ and $N=2\Omega+(n_p-n_h)$, 
$\dket{m}$ can be also specified by $\dket{\Omega,N;s}$. 
Then, the state $\dket{\Omega,N;ss_0}$ is given as 
\beq\label{4-14}
& &\dket{\Omega,N;ss_0}=({\hat S}_+(s_ps_h))^{s+s_0}\dket{\Omega,N;s} \ . 
\eeq
The above is the Holstein-Primakoff boson realization 
of the Lipkin model including the case $n_p\neq n_h$.

\section{The Lipkin model in the coupling of two kinds of the $su(2)$-spin}

Needless to say, the Lipkin model obeys the $su(2)$-algebra. 
The conventional form can be treated in terms of one kind of the $su(2)$-spin. 
But, the present Lipkin model, which we called the unconventional form, 
is treated in terms of the addition of two kinds of the $su(2)$-spins. 
This will be later shown. 
The aim of this section is to give a possible interpretation of the 
present Lipkin model in the frame of the coupling of two kinds of 
the $su(2)$-spins.

First, we discuss the case of the Schwinger representation. 
The form (\ref{3-1}) can be re-expressed in the following form: 
\beq\label{5-1}
& &{\hat S}_{\pm,0}={\hat S}_{\pm,0}(p)+{\hat S}_{\pm,0}(h) \ , 
\qquad\qquad\qquad\qquad\qquad\qquad
\qquad\qquad\ 
\eeq
\vspace{-0.5cm}
\bsub\label{5-2}
\beq
& &{\hat S}_+(p)={\hat a}_p^*{\hat b}_h\ , \qquad
{\hat S}_-(p)={\hat b}_h^*{\hat a}_p \ , \qquad
{\hat S}_0(p)=\frac{1}{2}({\hat a}_p^*{\hat a}_p-{\hat b}_h^*{\hat b}_h) \ , 
\label{5-2a}\\
& &{\hat S}_+(h)=-{\hat a}_h^*{\hat b}_p\ , \quad
{\hat S}_-(h)=-{\hat b}_p^*{\hat a}_h \ , \quad
{\hat S}_0(h)=\frac{1}{2}({\hat a}_h^*{\hat a}_h-{\hat b}_p^*{\hat b}_p) \ . 
\label{5-2b}
\eeq
\esub
We can see that the set of the generators $({\hat S}_{\pm,0})$ forms
simple sum of the two sets of the $su(2)$-generators, each of which is 
identical with the form presented in \S 2. 
Therefore, our problem is reduced to the addition of the $su(2)$-spins. 
The coupling scheme in the Schwinger representation has been 
formulated in detail by the present authors in Ref.\citen{13} 
and we copy some formulae from Ref.\citen{13}. 
Of course, the notations are changed from the original to the 
present ones. 
The eigenstate of ${\hat {\mib S}}^2$ and ${\hat S}_0$ with the 
eigenvalues $s(s+1)$ and $s_0$, respectively, is given in Ref.\citen{13}: 
\beq\label{5-3}
& &\ket{s_ps_h;ss_0}=({\hat T}_+)^{s_p+s_h-s}({\hat S}_+)^{s+s_0}
\ket{s_ps_h;s} \ , \nonumber\\
& &\ket{s_ps_h;s}=({\hat b}_p^*)^{s_h-s_p+s}({\hat b}_h^*)^{s-s_h+s_p}
\ket{0} \ .
\eeq
The exponents of ${\hat T}_+$, ${\hat b}_p^*$ and ${\hat b}_h^*$ should be 
positive and we have well known rule:
\beq\label{5-4}
& &|s_p-s_h|\leq s \leq s_p+s_h \ .
\eeq
Of course, the eigenvalues of ${\wtilde {\mib S}}(p)^2$ and 
${\wtilde {\mib S}}(h)^2$ are given by 
$s_p(s_p+1)$ and $s_h(s_h+1)$, respectively. 
The minimum weight state $\ket{m}$ is expressed as 
\beq\label{5-5}
& &\ket{m}=\ket{s_ps_h;s} \ . 
\eeq
We pay a special attention to the case 
\beq\label{5-6}
s_p+s_h=s\ .
\eeq
In this case, $\ket{m}$ is reduced to 
\beq\label{5-7}
& &\ket{s_ps_h;s=s_p+s_h}=({\hat b}_p^*)^{2s_h}({\hat b}_h^*)^{2s_p}
\ket{0} \ . 
\eeq
Further, we have 
\beq\label{5-8}
& &\ket{s_ps_h;s=s_p+s_h,s_0}=({\hat S}_+)^{s+s_0}\ket{s_ps_h;s} \ . 
\eeq
Since the state (\ref{5-8}) does not contain ${\hat T}_+$, the state 
(\ref{5-8}) satisfies 
\beq\label{5-9}
& &{\hat T}_-\ket{s_ps_h;s=s_p+s_h}=0 \ . 
\eeq
Therefore, the state investigated in \S 3 is nothing but the state 
(\ref{5-8}) with 
\beq\label{5-10}
& &s_p=\Omega-\frac{1}{2}(n_p+n_h)-\frac{1}{2}|n_p-n_h|\ , \qquad
s_h=\frac{1}{2}|n_p-n_h| \ . 
\eeq
From the above argument, we can learn that the operation of ${\hat T}_+$ 
is necessary for obtaining the states with $|s_p-s_h|\leq s < s_p+s_h$. 
However, these states do not have any counterparts to the original 
fermion states.

Next, we
treat the case of the Holstein-Primakoff representation. 
According to the authors' knowledge, there is no precedent for 
the formalism in which the coupling sheme of two $su(2)$-spins 
is treated in the frame of the Holstein-Primakoff representation. 
Therefore, newly we have to present the idea. 
In the same way as the previous case, 
${\hat S}_{\pm,0}(s_ps_h)$ can be decomposed into the following form:
\beq\label{5-11}
& &{\hat S}_{\pm,0}(s_ps_h)={\hat S}_{\pm,0}^{(p)}(s_p)
+{\hat S}_{\pm,0}^{(h)}(s_h) \ , 
\qquad\qquad\qquad\qquad\qquad\qquad\qquad\ \ 
\eeq
\vspace{-0.5cm}
\setcounter{equation}{10}
\bsub
\beq
& &{\hat S}_+^{(p)}(s_p)={\hat A}_p^*\sqrt{2s_p-{\hat A}_p^*{\hat A}_p}\ , 
\qquad
{\hat S}_-^{(p)}(s_p)=\sqrt{2s_p-{\hat A}_p^*{\hat A}_p}\ {\hat A}_p\ , 
\nonumber\\
& &{\hat S}_0^{(p)}(s_p)={\hat A}_p^*{\hat A}_p-s_p \ ,
\label{5-11a}\\
& &{\hat S}_+^{(h)}(s_h)=-{\hat A}_h^*\sqrt{2s_h-{\hat A}_h^*{\hat A}_h}\ , 
\qquad
{\hat S}_-^{(h)}(s_h)=-\sqrt{2s_h-{\hat A}_h^*{\hat A}_h}\ {\hat A}_h\ , 
\nonumber\\
& &{\hat S}_0^{(h)}(s_h)={\hat A}_h^*{\hat A}_h-s_h \ .
\label{5-11b}
\eeq
\esub
Our final goal is to find the eigenstate $\dket{s_ps_h;ss_0}$ in our 
present case, which is expressed as 
\beq\label{5-12}
& &\dket{s_ps_h;ss_0}=({\hat S}_+(s_ps_h))^{s+s_0}\dket{s_ps_h;s} \ .
\eeq
Here, $\dket{s_ps_h;s}$ denotes the minimum weight state: 
\beq\label{5-13}
& &{\hat S}_-\dket{s_ps_h;s}=0 \ , \qquad
{\hat S}_0\dket{s_ps_h;s}=-s\dket{s_ps_h;s} \ . 
\eeq
A possible choice of $\dket{s_ps_h;s}$ is given in the case 
$s=s_p+s_h$: 
In this case,\break
$\dket{s_ps_h;s(=s_p+s_h)}$ is the vacuum of ${\hat A}_p$ and 
${\hat A}_h$: 
\beq\label{5-14}
& &\dket{s_ps_h;s(=s_p+s_h)}=\dket{0} \ , \qquad
{\hat A}_p\dket{0}={\hat A}_h\dket{0}=0\ .
\eeq

Then, let us consider the minimum weight states in other cases. 
For this purpose, we introduce the following operators: 
\beq
& &{\hat S}_-(s_ps_h;k)=\sqrt{2s_p-k+1-{\hat A}_p^*{\hat A}_p}\ {\hat A}_p 
-\sqrt{2s_h-k+1-{\hat A}_h^*{\hat A}_h}\ {\hat A}_h  \ , 
\label{5-15}\\
& &{\hat R}_+(s_ps_h;k)={\hat A}_p^*\sqrt{2s_h-k+1-{\hat A}_h^*{\hat A}_h}
+{\hat A}_h^*\sqrt{2s_p-k+1-{\hat A}_p^*{\hat A}_p} \ , 
\label{5-16}\\
& &k=1,\ 2,\cdots ,\ k_m\ (=s_p+s_h-|s_p-s_h|) \ . 
\label{5-17}
\eeq
It should be noted that the case $k=1$ gives us 
\beq\label{5-18}
& &{\hat S}_-(s_ps_h;1)={\hat S}_-(s_ps_h)\ .
\eeq
The operators ${\hat S}_-(s_ps_h;k)$ and ${\hat R}_+(s_ps_h;k)$ are 
defined under the condition 
\beq\label{5-19}
& &2s_p-k \geq 0 \ , \qquad
2s_h-k \geq 0 \ , \quad 
{\rm i.e.,}\quad k\leq s_p+s_h-|s_p-s_h| \ .
\eeq
Therefore, the maximum value of $k$ is given as $k_m$ shown in the 
relation (\ref{5-17}). 
Direct calculation gives us that ${\hat S}_-(s_ps_h;k)$ and 
${\hat R}_+(s_ps_h;k)$ obey the relation 
\beq
& &{\hat S}_-(s_ps_h;k){\hat R}_+(s_ps_h;k)={\hat R}_+(s_ps_h;k)
{\hat S}_-(s_ps_h;k+1) \ , 
\label{5-20}\\
& &[\ {\hat S}_0(s_ps_h) \ , \ {\hat R}_+(s_ps_h;k)\ ]
={\hat R}_+(s_ps_h;k)\ . 
\label{5-21}
\eeq
With the use of ${\hat R}_+(s_ps_h;s)$, we can construct the minimum weight 
states in the form
\bsub\label{5-22}
\beq
& &\dket{s_ps_h;s(=s_p+s_h)}=\dket{0} \ , 
\label{5-22a}\\
& &\dket{s_ps_h;s}={\hat R}_+(s_ps_h;1){\hat R}_+(s_ps_h;2)\cdots
{\hat R}_+(s_ps_h;k(=s_p+s_h-s)\dket{0} \ , \nonumber\\
{\rm for}\ \ & &s=1,\ 2,\cdots, \ k_m(=s_p+s_h-|s_p-s_h|) \ .
\label{5-22b}
\eeq
\esub
The case (\ref{5-22a}) is self-evident. 
The proof for the case (\ref{5-22b}) which is the minimum weight state 
is performed in the following way: 
With the successive use of the condition (\ref{5-20}), we have 
\bsub\label{5-23}
\beq
& &{\hat S}_-(s_ps_h)\dket{s_ps_h;s} \nonumber\\
&=&{\hat S}_-(s_ps_h;1){\hat R}_+(s_ps_h;1){\hat R}_+(s_ps_h;2)\cdots 
{\hat R}_+(s_ps_h;k)\dket{0} \nonumber\\
&=&{\hat R}_+(s_ps_h;1){\hat R}_+(s_ps_h;2)\cdots 
{\hat R}_+(s_ps_h;k){\hat S}_-(s_ps_h;k+1)\dket{0}\ , 
\nonumber\\
& &\qquad\qquad\qquad\qquad\qquad\qquad\qquad\qquad\qquad
(|s_p-s_h|<s<s_p+s_h) 
\label{5-23a}\\
& &{\hat S}_-(s_ps_h)\dket{s_ps_h;s} \nonumber\\
&=&{\hat R}_+(s_ps_h;1){\hat R}_+(s_ps_h;2)\cdots {\hat R}_+(s_ps_h;k-1)
{\hat S}_-(s_ps_h;k){\hat R}_+(s_ps_h;k)\dket{0} \ . 
\nonumber\\
& &\qquad\qquad\qquad\qquad\qquad\qquad\qquad\qquad\qquad
\qquad\qquad    
(s=|s_p-s_h|) 
\label{5-23b}
\eeq
\esub
For the case (\ref{5-23a}), ${\hat S}_-(s_ps_h;k+1)\dket{0}=0$ and 
for the case (\ref{5-23b}),\break
${\hat S}_-(s_ps_h;k){\hat R}_+(s_ps_h;k)\dket{0}=0$. 
The reason why we treated separately comes from the condition (\ref{5-19}). 
Operation ${\hat S}_0(s_ps_h)$ on $\dket{s_ps_h;s}$ gives us 
\beq\label{5-24}
{\hat S}_0(s_ps_h)\dket{s_ps_h;s}
&=&(k-(s_p+s_h))\dket{s_ps_h;s} \nonumber\\
&=&-s\dket{s_ps_h;s} \ . 
\eeq
Here, we use the condition (\ref{5-21}) and $k=s_p+s_h-s$. 
Thus, we could complete the coupling scheme of two kinds of the 
$su(2)$-spin. 
But, in the same meaning as the case of the Schwinger representation, 
the case $s=s_p+s_h$ is the counterpart of the Lipkin model.



\section{The simplest approximate diagonalization of the Hamiltonian 
of the Lipkin model}

In our present boson representation, we can diagonalize the Hamiltonian of the Lipkin model exactly. 
But, it may be important to show an approximate solution on the 
same level as that obtained in the conventional random phase approximation. 
The Holstein-Primakoff representation may be suitable for this aim, 
because generally the operators describing the system under investigation are 
expressed in terms of the power series for the bosons playing a role of the 
fluctuations around the equilibrium. 

The present diagonalization is based on the following relation: 
\beq\label{6-1}
& &{\hat S}_+(s_p s_h)\approx {\hat A}_p^*\sqrt{2s_p}
-{\hat A}_h^*\sqrt{2s_h} \ , \nonumber\\
& &{\hat S}_-(s_p s_h)\approx \sqrt{2s_p}{\hat A}_p
-\sqrt{2s_h}{\hat A}_h \ , \nonumber\\
& &{\hat S}_0(s_p s_h)={\hat A}_p^*{\hat A}_p+
{\hat A}_h^*{\hat A}_h-s \ . \quad (s=s_p+s_h)
\eeq
The approximation (\ref{6-1}) indicates that the effects of 
${\hat A}_p^*{\hat A}_p$ and ${\hat A}_h^*{\hat A}_h$ are 
neglisibly small compared with $2s_p$ and $2s_h$, respectively, 
in the square root of the relation (\ref{4-11}): 
\beq\label{6-2}
& &\sqrt{2s_p-{\hat A}_p^*{\hat A}_p}\approx \sqrt{2s_p} \ , \qquad
\sqrt{2s_h-{\hat A}_h^*{\hat A}_h}\approx \sqrt{2s_h} \ .
\eeq
The relation (\ref{6-2}) shows that the existence of the equilibrium is presuposed. 
For the relation (\ref{6-1}), we introduce the boson operators 
\bsub\label{6-3}
\beq
& &{\hat B}^*=\sqrt{\frac{s_p}{s}}{\hat A}_p^*
-\sqrt{\frac{s_h}{s}}{\hat A}_h^*\ , \qquad
{\hat B}=\sqrt{\frac{s_p}{s}}{\hat A}_p
-\sqrt{\frac{s_h}{s}}{\hat A}_h \ , \nonumber\\
& &{\hat C}^*=\sqrt{\frac{s_h}{s}}{\hat A}_p^*
+\sqrt{\frac{s_p}{s}}{\hat A}_h^*\ , \qquad
{\hat C}=\sqrt{\frac{s_h}{s}}{\hat A}_p
+\sqrt{\frac{s_p}{s}}{\hat A}_h \ . 
\label{6-3a}
\eeq
Conversely, 
\beq
& &{\hat A}_p^*=\sqrt{\frac{s_p}{s}}{\hat B}^*
+\sqrt{\frac{s_h}{s}}{\hat C}^*\ , \qquad
{\hat A}_p=\sqrt{\frac{s_p}{s}}{\hat B}
+\sqrt{\frac{s_h}{s}}{\hat C} \ , \nonumber\\
& &{\hat A}_h^*=-\sqrt{\frac{s_h}{s}}{\hat B}^*
+\sqrt{\frac{s_p}{s}}{\hat C}^*\ , \qquad
{\hat A}_h=-\sqrt{\frac{s_h}{s}}{\hat B}
+\sqrt{\frac{s_p}{s}}{\hat C} \ . 
\label{6-3b}
\eeq
\esub
Then, we have 
\beq\label{6-4}
& &{\hat S}_+(s_ps_h)=\sqrt{2s}{\hat B}^* \ , \qquad
{\hat S}_-(s_ps_h)=\sqrt{2s}{\hat B} \ , 
\nonumber\\
& &{\hat S}_0(s_ps_h)={\hat B}^*{\hat B}+{\hat C}^*{\hat C}-s\ . 
\eeq
Hereafter, the symbol $\approx$ is replaced with the equal sign. 

\begin{figure}[b]
\begin{center}
\includegraphics[height=4.8cm]{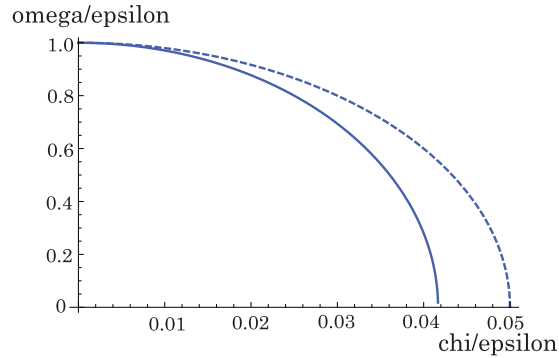}
\caption{$\omega/\epsilon$ for $n_p=n_h=0$ (solid curve) and $n_p=n_h=1$ (dashed curve)
with $N=12$ are shown.
}
\label{fig:1}
\end{center}
\end{figure}

\begin{figure}[t]
\begin{center}
\includegraphics[height=4.5cm]{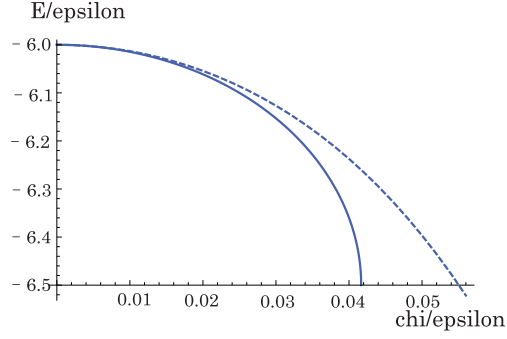}
\caption{The ground state energy for $n_p=n_h=0$ (solid curve) and exact ground state 
energy (dashed curve) are shown in the case of the closed shell system with $N=12$.
}
\label{fig:2}
\end{center}
\end{figure}
%
\begin{figure}[t]
\begin{center}
\includegraphics[height=4.5cm]{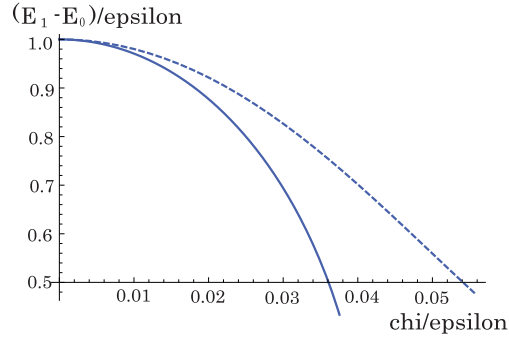}
\caption{The energy difference between ground state and the first excited state for $n_p=n_h=0$ 
(solid curve) and exact results (dashed curve) are shown in the case of the closed shell system 
with $N=12$.
}
\label{fig:3}
\end{center}
\end{figure}

Under the relation (\ref{6-3}), the Hamiltonian (\ref{2-4}) is approximated to 
\beq\label{6-5}
& &{\hat H}=\epsilon ({\hat B}^*{\hat B}+{\hat C}^*{\hat C}-s)
-2\chi s({\hat B}^{*2}+{\hat B}^2) \ . 
\eeq
The Hamiltonian (\ref{6-5}) is easily diagonalized in the form 
\beq
& &{\hat H}=E_0+\epsilon {\hat C}^*{\hat C}+\omega(s) {\hat D}^*{\hat D} \ , 
\label{6-6}\\
& &E_0=-\frac{1}{2}(\epsilon -\omega(s))-\epsilon s \ , 
\label{6-7}\\
& &\omega(s)=\sqrt{\epsilon^2-(4\chi s)^2} \ . 
\label{6-8}
\eeq
The operator $({\hat D}^*,{\hat D})$ is also boson operator defined as 
\beq\label{6-9}
& &{\hat D}^*=\sqrt{\frac{\epsilon +\omega(s)}{2\omega(s)}}{\hat B}^*
-\sqrt{\frac{\epsilon-\omega(s)}{2\omega(s)}}{\hat B} \ , \quad
{\hat D}=\sqrt{\frac{\epsilon +\omega(s)}{2\omega(s)}}{\hat B}
-\sqrt{\frac{\epsilon-\omega(s)}{2\omega(s)}}{\hat B}^* \ . \quad
\eeq
The above result is reduced to that based on the random phase approximation for the 
closed shell system, if $s$ is equal to $\Omega$, 
i.e., $n_p=n_h=0$. 
The eigenstate and the eigenvalue are given as follows: 
\bsub\label{6-10}
\beq
& &\dket{\lambda \mu}=({\hat D}^*)^{\lambda}({\hat C}^*)^{\mu}\dket{\phi} \ , 
\qquad 
{\hat D}\dket{\phi}={\hat C}\dket{\phi}=0 \ , 
\label{6-10a}\\
& &E_{\lambda\mu}=E_0+\lambda\omega(s) +\mu\epsilon \ . 
\label{6-10b}
\eeq
\esub
The above is the result of the diagonalization of the boson Hamiltonian 
(\ref{6-5}).

\begin{figure}[t]
\begin{center}
\includegraphics[height=5.1cm]{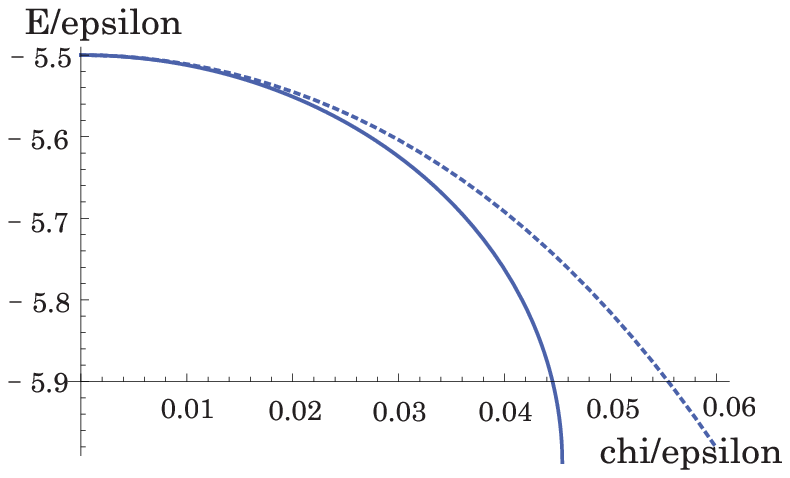}
\caption{The ground state energy for $n_p=0,\ n_h=1$ (solid curve) and exact ground state 
energy (dashed curve) are shown in the case of the open shell system with $N=11$ and 
$\Omega=6$.
}
\label{fig:4}
\end{center}
\end{figure}
%
%
\begin{figure}[t]
\begin{center}
\includegraphics[height=5.2cm]{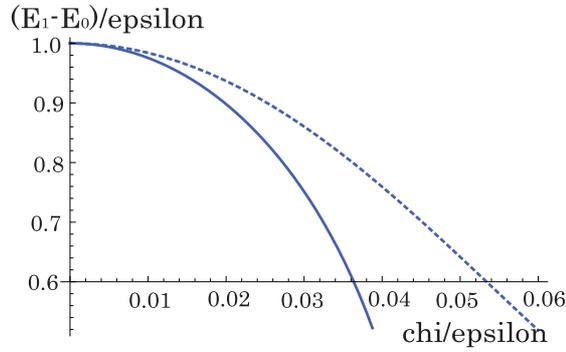}
\caption{The energy difference between ground state and the first excited state for $n_p=0,\ 
n_h=1$ (solid curve) and exact results (dashed curve) are shown in the case of the open shell system 
with $N=11$ and $\Omega=6$.
}
\label{fig:5}
\end{center}
\end{figure}

However, it must be noticed that the result (\ref{6-10}) contains a problem 
to be investigated. 
The result (\ref{6-10}) is derived for the Hamiltonian (\ref{6-5}) 
and it must be checked if the result (\ref{6-10}) is derived for the 
Hamiltonian (\ref{2-4}) or not. 
For this problem, we note the operator ${\hat R}_+(s_ps_h;k)$ 
defined in the relation (\ref{5-16}). 
Under the same sprit as that for the relation (\ref{6-2}), 
${\hat R}_+(s_ps_h;k)$ may be approximated as 
\beq\label{6-11}
& &{\hat R}_+(s_ps_h;k)\approx 
{\hat A}_p^*\sqrt{2s_h}+{\hat A}_h^*\sqrt{2s_p}=\sqrt{2s}{\hat C}^* \ .
\eeq
Therefore, under the present approximation, successive operation of 
${\hat R}_+(s_ps_h;k)$ may be equivalent to that of ${\hat C}^*$. 
On the other hand, the operation of ${\hat S}_+(s_ps_h)$ may be equivalent 
to that of  ${\hat B}^*$. 
Any state obtained by operation of ${\hat C}^*$ does not have the 
counterpart in the original fermion space. 
Thus, we have the following conclusion: 
Only the part related to $\mu=0$ should be selected such as 
\beq\label{6-12}
& &\dket{\lambda \mu=0}=({\hat D}^*)^{\lambda}\dket{\phi} \ , 
\qquad
E_{\lambda \mu=0}=E_0+\lambda \omega(s) \ . 
\eeq
Of course, the above result is applicable to the region 
\beq\label{6-13}
& &4\chi s \leq \epsilon \ . 
\eeq
The phase transition occurs at $\chi=\epsilon/4s$. 
The relation (\ref{6-13}) shows that the force strength $\chi$ 
at the phase transition point obtained in the case $s < \Omega$ 
becomes larger than $\chi$ obtained in the conventional 
random phase approximation for the case $s=\Omega$. 
From the above consideration, we can understand that 
our present approximation is a natural generalization from the 
conventional random phase approximation. 
Next, we will show some numerical results.

\begin{figure}[t]
\begin{center}
\includegraphics[height=5.0cm]{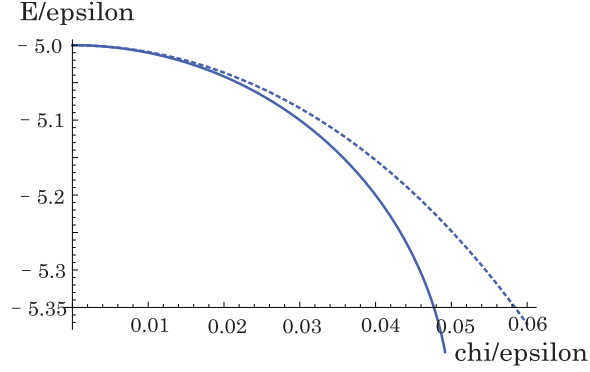}
\caption{The ground state energy for $n_p=0,\ n_h=2$ (solid curve) and exact ground state 
energy (dashed curve) are shown in the case of the open shell system with $N=10$ and 
$\Omega=6$.
}
\label{fig:6}
\end{center}
\end{figure}

\begin{figure}[t]
\begin{center}
\includegraphics[height=5.2cm]{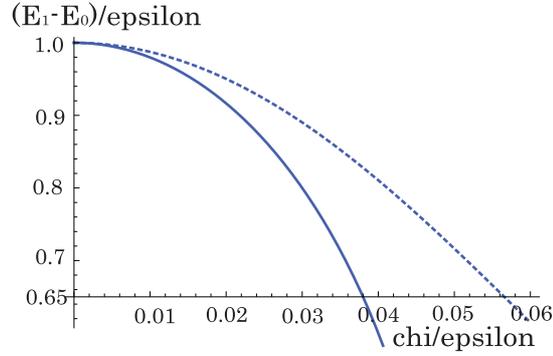}
\caption{The energy difference between ground state and the first excited state for $n_p=0,\ 
n_h=2$ (solid curve) and exact results (dashed curve) are shown in the case of the open shell system 
with $N=10$ and $\Omega=6$.
}
\label{fig:7}
\end{center}
\end{figure}

In Fig.\ref{fig:1}, $\omega/\epsilon$ in Eq.(\ref{6-8}) is depicted as a function of 
the force strength $\chi$ devided by $\epsilon$ in the cases 
$n_p=n_h=0$ (solid curve) and $n_p=n_h=1$ (dashed curve), respectively, with the fermion number $N=12$. 

First, let us consider the closed shell system with $n_p=n_h=0$.  
In Fig.\ref{fig:2}, the ground state energy for $n_p=n_h=0$ (solid curve) is 
compared with the exact ground state energy (dashed curve) as a function of the force strength $\chi$ 
in the case $N=12$ and 
$\Omega=6$, 
where all quantities are scaled by the single particle energy $\epsilon$. 
It is shown that this approximation is rather good in the region with small force strength 
compared with the phase transition point $\chi=\epsilon/4s$. 
In Fig.\ref{fig:3}, the energy difference between the first excited state and the ground state 
is depicted compared with the exact energy difference. 
It is seen that the goodness of the approximation is similar to that for the ground state energy.

\begin{figure}[t]
\begin{center}
\includegraphics[height=5.0cm]{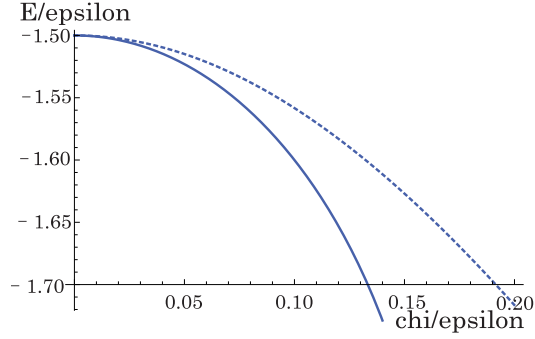}
\caption{The ground state energy for $n_p=0,\ n_h=9$ (solid curve) and exact ground state 
energy (dashed curve) are shown in the case of the open shell system with $N=3$ and 
$\Omega=6$.
}
\label{fig:10}
\end{center}
\end{figure}

\begin{figure}[t]
\begin{center}
\includegraphics[height=4.5cm]{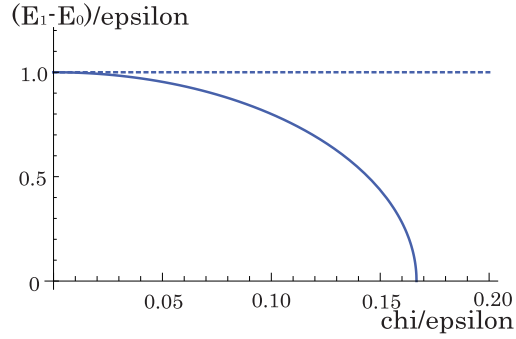}
\caption{The energy difference between ground state and the first excited state for $n_p=0,\ 
n_h=9$ (solid curve) and exact results (dashed curve) are shown in the case of the open shell system 
with $N=3$ and $\Omega=6$.
}
\label{fig:11}
\end{center}
\end{figure}

Next, let us consider the open shell system in which 
we take $n_p=0$ and $n_h \neq 0$. 
Figure \ref{fig:4} shows the ground state energy (solid curve) in $\Omega=6$ 
and $n_p=0$ and $n_h=1$, which leads to $N=11$, 
and the exact energy eigenvalue (dashed curve). 
In our framework, it is shown that we can describe the open shell system well 
as for the ground state energy.  
In Fig.\ref{fig:5}, the energy difference between the first excited state 
and the ground state is depicted with the same parameter set as that in Fig.\ref{fig:4}. 

The similar results are derived in the case $n_p=0$ and $n_h=2$ with $\Omega=6$  
which leads to $N=10$. 
Figures \ref{fig:6} and \ref{fig:7} show the calculated results of the 
ground state energies and 
the enegy differences between the first excited state and the ground state, 
respectively, for our approximated treatment and the exact diagonalization.



\begin{figure}[t]
\begin{center}
\includegraphics[height=5.0cm]{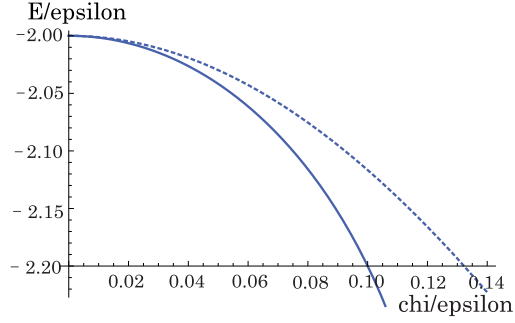}
\caption{The ground state energy for $n_p=0,\ n_h=8$ (solid curve) and exact ground state 
energy (dashed curve) are shown in the case of the open shell system with $N=4$ and 
$\Omega=6$.
}
\label{fig:12}
\end{center}
\end{figure}

\begin{figure}[t]
\begin{center}
\includegraphics[height=5.0cm]{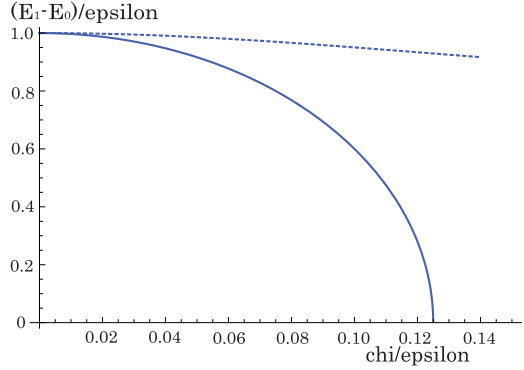}
\caption{The energy difference between ground state and the first excited state for $n_p=0,\ 
n_h=8$ (solid curve) and exact results (dashed curve) are shown in the case of the open shell system 
with $N=4$ and $\Omega=6$.
}
\label{fig:13}
\end{center}
\end{figure}

Further, let us consider the cases with small particle number in the open shell system. 
Figures \ref{fig:10} and \ref{fig:11} show the ground state energy and the energy difference 
between the first excited and ground state, respectively, in the case $N=3$ 
with $\Omega=6$, $n_p=0$ and $n_h=9$. 
The energy difference between the first excited state and the ground state 
does not depend on the force strength $\chi$ in the exact result. 
Except for this situation, the behavior is similar to the case with large $N$ in 
Figs.\ref{fig:2}$\sim$\ref{fig:7}. 
Also, in the case $N=4$ with $\Omega=6$, $n_p=0$ and $n_h=8$, the similar results 
are obtained as is shown in Figs.\ref{fig:12} and \ref{fig:13}.

In cnclusion, we can treat the open shell system by using of our unconventional 
boson realization method developed in this paper, adding to the case of 
the closed shell system. 
As a result, in spite of the approximation in Eq.(\ref{6-2}), 
in which it is assumed that the effects expressed by the power series of ${\hat A}_p^*{\hat A}_p$ and 
${\hat A}_h^*{\hat A}_h$ appearing in ${\hat S}_{\pm}(s_ps_h)$ are small compared with 
$2s_p$ and $2s_h$, respectively,  
the obtained results are rather good, especially, in the region 
with small force strength, while the approximation is not so good when 
the force strength approaches to the transition point.

\section{The isoscalar pairing model}

In addition to the Lipkin model, we know a many-fermion system consisting of the two 
single-particle levels and obeying the $su(2)$-algebra: 
The isoscalar pairing model. 
In this model, the two single-particle levels, which we call the p- and the n-level, 
are occupied by protons and neutrons, respectively. 
Of course, the degeneracies are the same as each other: 
$\Omega=j+1/2$ ($j$ : half-integer). 
Building block of this model is the proton-neutron pair coupled in the isoscalar 
type, which obeys the $su(2)$-algebra. 
Therefore, the isoscalar model is in a near relation to the Lipkin model. 
In this sense, it may be ineteresting to investigate both models comparatively. 
In this connection, the isovector pairing model obeys the $so(5)$-algebra and 
if both pairing models are combined with each other, we have the $su(4)$-algebra. 
The passage has been discussed in the high temperature superconductivity.\cite{9,10}

We denote the proton and the neutron operator as $(p_m,p_m^*)$ and $(n_m,n_m^*)$, 
respectively. 
Here, of course, $m=-j,\ -j+1, \cdots ,j-1,\ j$. 
In this model, we can introduce a set of the operators $({\wtilde \sigma}_{\pm,0})$ 
defined as 
\beq\label{7-1}
& &{\wtilde \sigma}_+=\sum_m \theta(m) p_m^*n_{\wtilde m}^* \ , \qquad
{\wtilde \sigma}_-=\sum_m \theta(m) n_{\wtilde m} p_{m} \ , 
\nonumber\\
& &{\wtilde \sigma}_0=\frac{1}{2}\sum_m (p_m^*p_m+n_m^*n_m)-\Omega \ . 
\eeq
Here, $\theta(m)=m/|m|$, i.e., $\theta(m)=1$ for $m>0$ and $\theta(m)=-1$ 
for $m<0$. 
The operator ${\wtilde \sigma}_+$ is expressed in a form of a certain linear 
combination of $(p^*n^*)_{J={\rm odd},M=0}$. 
If $\theta(m)=1$ for all $m$, ${\wtilde \sigma}_+=(p^*n^*)_{J=0,M=0}$ and 
in this case, including $(p^*p^*)_{J=0,M=0}$, $(n^*n^*)_{J=0,M=0}$, they form the 
$so(5)$-algebra.
It is easily verified that $({\wtilde \sigma}_{\pm,0})$ defined in the 
relation (\ref{7-1}) obeys 
\beq\label{7-2}
& &[\ {\wtilde \sigma}_+\ , \ {\wtilde \sigma}_-\ ]=2{\wtilde \sigma}_0 \ , \qquad
[\ {\wtilde \sigma}_0\ , \ {\wtilde \sigma}_{\pm}\ ]=\pm{\wtilde \sigma}_{\pm} \ . 
\eeq
Further, we have 
\beq\label{7-3}
& &[\ {\rm any\ of}\ ({\wtilde \tau}_{\pm,0})\ , \ {\rm any\ of}\ ({\wtilde \sigma}_{\pm,0})\ ]=0 \ .
\eeq
Here, $({\wtilde \tau}_{\pm,0})$ denotes isospin operator: 
\beq\label{7-4}
& &{\wtilde \tau}_+=\sum_m p_m^*n_m\ , \qquad
{\wtilde \tau}_-=\sum_m n_m^*p_m\ , \qquad
{\wtilde \tau}_0=\frac{1}{2}\sum_m (p_m^*p_m -n_m^*n_m)\ .
\eeq
The relation (\ref{7-3}) tells us that $({\wtilde \sigma}_{\pm,0})$ is isoscalar. 
Different from the case of the Lipkin model, the operator ${\wtilde \sigma}_0$ 
is expressed in terms of the total nucleon number ${\wtilde N}$: 
\beq\label{7-5}
{\wtilde N}={\wtilde N}_p+{\wtilde N}_n \ , \qquad
{\wtilde N}_p=\sum_m p_m^*p_m \ , \qquad
{\wtilde N}_n=\sum_m n_m^*n_m \ . 
\eeq
This fact is important. 
It gives us  a possible boson realization of the $su(2)$-algebra which is different 
from the Lipkin model treated in \S\S 3 and 4. 
The relations (\ref{7-1}), (\ref{7-4}) and (\ref{7-5}) give us 
\beq\label{7-6}
& &{\wtilde N}_p=\Omega+{\wtilde \sigma}_0+{\wtilde \tau}_0 \ , \qquad
{\wtilde N}_n=\Omega+{\wtilde \sigma}_0-{\wtilde \tau}_0 \ .
\eeq

By replacing the index $h$ with $n$ in the relation (\ref{3-1}), we postulate the counterpart 
of ${\wtilde \sigma}_{\pm,0}$, which we denote ${\hat \sigma}_{\pm,0}$, in the following 
form: 
\beq\label{7-7}
& &{\hat \sigma}_+={\hat a}_p^*{\hat b}_n-{\hat a}_n^*{\hat b}_p \ , \qquad
{\hat \sigma}_-={\hat b}_n^*{\hat a}_p-{\hat b}_p^*{\hat a}_n \ , \nonumber\\
& &{\hat \sigma}_0=\frac{1}{2}\left[({\hat a}_p^*{\hat a}_p+{\hat a}_n^*{\hat a}_n)
-({\hat b}_p^*{\hat b}_p+{\hat b}_n^*{\hat b}_n)\right] \ .
\eeq
The $su(1,1)$-generators are given as 
\beq\label{7-8}
& &{\hat T}_+={\hat a}_p^*{\hat b}_p^*+{\hat a}_n^*{\hat b}_n^* \ , \qquad
{\hat T}_-={\hat b}_p{\hat a}_p+{\hat b}_n{\hat a}_n \ , \nonumber\\
& &{\hat T}_0=\frac{1}{2}\left[({\hat a}_p^*{\hat a}_p+{\hat a}_n^*{\hat a}_n)
+({\hat b}_p^*{\hat b}_p+{\hat b}_n^*{\hat b}_n)\right]+1 \ .
\eeq
Of course, the above expression comes from the relation (\ref{3-2}). 
Further, we postulate the following form for the counterpart of $({\wtilde \tau}_{\pm,0})$:
\beq\label{7-9}
& &{\hat \tau}_+={\hat a}_p^*{\hat a}_n-{\hat b}_p^*{\hat b}_n \ , \qquad
{\hat \tau}_-={\hat a}_n^*{\hat a}_p-{\hat b}_n^*{\hat b}_p \ , \nonumber\\
& &{\hat \tau}_0=\frac{1}{2}\left[({\hat a}_p^*{\hat a}_p-{\hat a}_n^*{\hat a}_n)
-({\hat b}_p^*{\hat b}_p-{\hat b}_n^*{\hat b}_n)\right] \ .
\eeq
The set $({\hat \tau}_{\pm,0})$ obeys the $su(2)$-algebra and commutes with 
$({\hat \sigma}_{\pm,0})$ and $({\hat T}_{\pm,0})$. 
The relation (\ref{7-6}) permitts us to set up the relation 
\beq\label{7-10}
& &{\wtilde N}_p \rightarrow {\hat N}_p={\hat \Omega}+{\hat \sigma}_0+{\hat \tau}_0 
={\hat \Omega}+{\hat a}_p^*{\hat a}_p-{\hat b}_p^*{\hat b}_p \ , \nonumber\\
& &{\wtilde N}_n \rightarrow {\hat N}_n={\hat \Omega}+{\hat \sigma}_0-{\hat \tau}_0 
={\hat \Omega}+{\hat a}_n^*{\hat a}_n-{\hat b}_n^*{\hat b}_n \ . 
\eeq
The operator ${\hat \Omega}$ is determined in the framework of the form
\beq\label{7-11}
& &{\hat \Omega}=x+\frac{y}{2}({\hat a}_p^*{\hat a}_p+{\hat a}_n^*{\hat a}_n
+{\hat b}_p^*{\hat b}_p+{\hat b}_n^*{\hat b}_n) \ . 
\eeq
Of course, $x$ and $y$ should be determined.

For the minimum weight state $\ket{m}$, we require the relations 
\beq
& &{\hat \sigma}_-\ket{m}={\hat T}_-\ket{m}=0 \ , 
\label{7-12}\\
& &{\hat N}_p\ket{m}=n_p\ket{m} \ , \qquad
{\hat N}_n\ket{m}=n_n\ket{m}\ , \qquad
{\hat \Omega}\ket{m}=\Omega\ket{m} \ . 
\label{7-13}
\eeq
The relations (\ref{7-12}) and (\ref{7-13}) gives us 
\beq
& &{\hat \Omega}=\frac{1}{2}(n_p+n_n)+\frac{1}{2}({\hat a}_p^*{\hat a}_p+{\hat a}_n^*{\hat a}_n
+{\hat b}_p^*{\hat b}_p+{\hat b}_n^*{\hat b}_n) \ , 
\label{7-14}\\
& &\ket{m}=({\hat b}_n^*)^{\Omega-n_n}({\hat b}_p^*)^{\Omega-n_p}\ket{0} \ . 
\label{7-15} 
\eeq
Then, we have 
\beq\label{7-16}
& &{\hat N}_p=\frac{1}{2}(n_p+n_n)+\frac{1}{2}
(3{\hat a}_p^*{\hat a}_p+{\hat a}_n^*{\hat a}_n-{\hat b}_p^*{\hat b}_p+{\hat b}_n^*{\hat b}_n) \ , 
\nonumber\\
& &{\hat N}_n=\frac{1}{2}(n_p+n_n)+\frac{1}{2}
({\hat a}_p^*{\hat a}_p+3{\hat a}_n^*{\hat a}_n+{\hat b}_p^*{\hat b}_p-{\hat b}_n^*{\hat b}_n) \ . 
\eeq
Namely, 
\beq\label{7-17}
& &{\hat N}=n_p+n_n+2({\hat a}_p^*{\hat a}_p+{\hat a}_n^*{\hat a}_n) \ .
\eeq
The state $\ket{m}$ is also the eigenstate for ${\hat \sigma}_0$ with the 
eigenvalue 
\beq\label{7-18}
& &\sigma=\Omega-\frac{1}{2}(n_p+n_n) \ . 
\eeq
We can see that ${\hat \Omega}$ and $\sigma$ are formally identical to those in the 
Lipkin model, but, ${\hat N}$ is different from that in the Lipkin model. 
It is nothing but the expectation.

Next task is to discuss the orthogonal set for the present model. 
The minimum weight state $\ket{m}$ shown in the form (\ref{7-5}) satisfies the relation 
\beq
& &{\hat \sigma}_0\ket{m}=-\sigma\ket{m} \ , \qquad
{\hat \tau}_0\ket{m}=\tau\ket{m} \ , \qquad
{\hat T}_0\ket{m}=T\ket{m} \ , 
\label{7-19}\\
& &\sigma=\Omega-\frac{1}{2}(n_p+n_n) \ , \qquad
\tau_0=\frac{1}{2}(n_p-n_n) \ , \qquad
T=\Omega+1 \ . 
\label{7-20}
\eeq
Further, for the Casimir operators ${\hat {\mib \sigma}}^2$, ${\hat {\mib \tau}}^2$ and 
${\hat {\mib T}}^2$, we have 
\beq\label{7-21}
& &{\hat {\mib \sigma}}^2\ket{m}={\hat {\mib \tau}}^2\ket{m}={\hat {\mib T}}^2\ket{m}
=\sigma(\sigma+1)\ket{m} \ . 
\eeq
Then, $\ket{m}$ can be rewritten in the form 
\beq\label{7-22}
& &\ket{\Omega;\sigma\tau_0}=({\hat b}_n^*{\hat b}_p)^{\sigma+\tau_0}
({\hat b}_p^*)^{2\sigma}\ket{0}=
({\hat \tau}_+)^{\sigma+\tau_0}({\hat b}_p^*)^{2\sigma}\ket{0} \ . 
\eeq
Of course, including the phase factor, the normalization is arbitrary. 
Therefore, the eigenstate of ${\hat \sigma}_0$ with the eigenvalue 
$\sigma_0$ is given in the form 
\beq\label{7-23}
& &\ket{\Omega;\sigma\sigma_0\tau_0}=({\hat \sigma}_+)^{\sigma+\sigma_0}
({\hat \tau}_+)^{\sigma+\tau_0}({\hat b}_p^*)^{2\sigma}\ket{0} \ . 
\eeq
Since $[{\hat \sigma}_+ , {\hat \tau}_+]=0$, the state $\ket{\Omega;\sigma\sigma_0\tau_0}$ 
can be expressed in various forms, for example, 
\beq\label{7-24}
& &\ket{\Omega;\sigma\sigma_0\tau_0}=({\hat \tau}_+)^{\sigma+\tau_0}({\hat \sigma}_+)^{\sigma+\sigma_0}
({\hat b}_p^*)^{2\sigma}\ket{0} \ . 
\eeq
With the use of the relation (\ref{7-17}), the eigenvalue of ${\hat N}$, $N$, for 
the state $\ket{\Omega;\sigma\sigma_0\tau_0}$ is expressed as 
\beq\label{7-25}
& &N=2\Omega-2\sigma_0 \ , \quad {\rm i.e.,}\quad
\sigma_0=\Omega-\frac{N}{2} \ . 
\eeq
Therefore, we have 
\beq\label{7-26}
\sigma=\Omega-\frac{1}{2}(n_p+n_n) \ , \qquad
\sigma_0=\Omega-\frac{N}{2}\ , \qquad
\tau_0=\frac{1}{2}(n_p-n_n) \ . 
\eeq
Since $n_p \geq 0$, $n_n \geq 0$, $-\sigma \leq \sigma_0 \leq \sigma$, $n_p$, $n_n$ and $N$ 
obey the inequality 
\beq\label{7-27}
& &0\leq n_p \leq \Omega\ , \qquad
0\leq n_n \leq \Omega\ , \qquad
n_p+n_n \leq N \leq 4\Omega-(n_p+n_n) \ .
\eeq
In this way, we know that $\ket{\Omega;\sigma\sigma_0\tau_0}$ is characterized by $\Omega$, 
$N$, $n_p$ and $n_n$ governed by the relation (\ref{7-27}). 
Of course, $n_p$ and $n_n$ determine the minimum weight state.

Finally, we will sketch the idea for constructing the Holstein-Primakoff 
boson realization. 
We introduce the operators ${\hat \sigma}_p$ and ${\hat \sigma}_n$ defined as 
\beq\label{7-28}
& &{\hat \sigma}_p=\frac{1}{2}({\hat a}_p^*{\hat a}_p+{\hat b}_p^*{\hat b}_p)\ , \qquad
{\hat \sigma}_n=\frac{1}{2}({\hat a}_n^*{\hat a}_n+{\hat b}_n^*{\hat b}_n)\ . 
\eeq
Since $[{\hat \sigma}_p , {\hat \sigma}_{\pm,0}]=[{\hat \sigma}_n,{\hat \sigma}_{\pm,0}]=0$, 
we have  
\beq\label{7-29}
& &{\hat \sigma}_p\ket{\Omega;\sigma\sigma_0\tau_0}=\sigma_p\ket{\Omega;\sigma\sigma_0\tau_0} \ , 
\qquad
\sigma_p=\frac{1}{2}(\Omega-n_p) \ , 
\nonumber\\
& &{\hat \sigma}_n\ket{\Omega;\sigma\sigma_0\tau_0}=\sigma_n\ket{\Omega;\sigma\sigma_0\tau_0} \ , 
\qquad
\sigma_n=\frac{1}{2}(\Omega-n_n) \ . 
\eeq
Therefore, we have 
\beq\label{7-30}
& &\sigma=\sigma_p+\sigma_n \ .
\eeq
The above consideration enables us to construct the Holstein-Primakoff 
boson realization in parallel with the Lipkin model.

\section{Concluding remark}

In this paper, concentrating on the $su(2)$-algebraic many-fermion theory, 
a new framework of the boson realization for the $su(2)$-algebra was formulated. 
New boson realization developed in this paper is called the unconventional boson realization.  
In order to be able to treat the open shell system in the $su(2)$-algebraic 
model, the Schwinger and Holstein-Primakoff boson realizations were formulated 
in the unconventional form. 

The unconventional boson realization method developed in this paper 
was applied to the two-level Lipkin model 
whose Hamiltonian can be expressed in terms of the generators of the $su(2)$-algebra.
One can deal with the closed shell system only by using the conventional 
boson realization such as the Schwinger and the Holstein-Primakoff boson realization. 
However, it was shown that, in our formalism with the unconventional boson realization, 
the open shell system with $N \neq 2\Omega$ can be described. 
As an concrete example,  
the ground state energy and the energy difference between the ground state and the first excited 
state were investigated by using the Holstein-Primakoff-type unconventinal boson realization 
under a certain approximation. 
This approximation corresponds to the random phase approximation describing the closed shell 
system. 
It was demonstrated that the calculated results in the region with small force strength are 
rather good in comparison with the exact results, 
while the approximated results are not so good when the force strength approaches the phase transition point. 
The behavior of the approximate solution mentioned above may be conjectured beforehand, 
because the approximation adopted in \S 6 is on the same level as that in the case 
of the closed shell system, i.e., the random phase approximation. 
In order to interpret the behavior near the phase transition point, various ideas have been 
proposed. 
We can apply these ideas to the present system, but, this investigation is a problem to be 
solved in future.

Further, adding to the two-level Lipkin model, we discussed the isoscalar pairing model in our 
unconventional boson realization comparatively. 
It may be interesting to extending this model to the isovector pairing model governed by 
the $so(5)$-algebra and, also, the combined model with the isoscalar and the isovector pairing model 
governed by the $su(4)$-algebra.

\section*{Acknowledgements} 
One of the authors (Y.T.) 
is partially supported by the Grants-in-Aid of the Scientific Research 
No.18540278 from the Ministry of Education, Culture, Sports, Science and 
Technology in Japan.



\end{document}